\documentstyle[12pt,aaspp4]{article}
\begin{document}

\newcommand{\up}[1]{\ifmmode^{\rm #1}\else$^{\rm #1}$\fi}
\newcommand{\zdot}{\makebox[0pt][l]{.}}
\newcommand{\upd}{\up{d}}
\newcommand{\uph}{\up{h}}
\newcommand{\upm}{\up{m}}
\newcommand{\ups}{\up{s}}
\newcommand{\arcd}{\ifmmode^{\circ}\else$^{\circ}$\fi}
\newcommand{\arcm}{\ifmmode{'}\else$'$\fi}
\newcommand{\arcs}{\ifmmode{''}\else$''$\fi}

\title{The Araucaria Project. A Wide-Field Photometric 
Survey for Cepheid Variables in NGC 3109
\footnote{Based on  observations obtained with the 1.3 m Warsaw 
telescope at Las Campanas Observatory
}
}
\author{Grzegorz Pietrzy{\'n}ski}
\affil{Universidad de Concepci{\'o}n, Departamento de Fisica, Astronomy
Group,
Casilla 160-C,
Concepci{\'o}n, Chile}
\affil{Warsaw University Observatory, Al. Ujazdowskie 4, 00-478, Warsaw,
Poland}
\authoremail{pietrzyn@hubble.cfm.udec.cl}
\author{Wolfgang Gieren}
\affil{Universidad de Concepci{\'o}n, Departamento de Fisica, Astronomy Group,
Casilla 160-C, Concepci{\'o}n, Chile}
\authoremail{wgieren@astro-udec.cl}
\author{Andrzej  Udalski}
\affil{Warsaw University Observatory, Al. Ujazdowskie 4, 00-478, Warsaw,
Poland}
\authoremail{udalski@astrouw.edu.pl}
\author{Igor Soszy{\'n}ski}
\affil{Universidad de Concepci{\'o}n, Departamento de Fisica, Astronomy Group, 
Casilla 160-C, Concepci{\'o}n, Chile}
\affil{Warsaw University Observatory, Al. Ujazdowskie 4, 00-478, Warsaw,
Poland}
\authoremail{soszynsk@astro-udec.cl}
\author{Fabio Bresolin}
\affil{Institute for Astronomy, University of Hawaii at Manoa, 2680 Woodlawn 
Drive, 
Honolulu HI 96822, USA}
\authoremail{bresolin@ifa.hawaii.edu}
\author{Rolf-Peter Kudritzki}
\affil{Institute for Astronomy, University of Hawaii at Manoa, 2680 Woodlawn 
Drive,
Honolulu HI 96822, USA}
\authoremail{kud@ifa.hawaii.edu}
\author{Ronald Mennickent}
\affil{Universidad de Concepci{\'o}n, Departamento de Fisica, Astronomy
Group, Casilla 160-C, Concepci{\'o}n, Chile}
\authoremail{rmennick@astro-udec.cl}
\author{Marcin Kubiak}
\affil{Warsaw University Observatory, Al. Ujazdowskie 4, 00-478, Warsaw,
Poland}
\authoremail{mk@astrouw.edu.pl}
\author{Micha{\l} Szyma{\'n}ski}
\affil{Warsaw University Observatory, Al. Ujazdowskie 4, 00-478, Warsaw,
Poland}
\authoremail{msz@astrouw.edu.pl}
\author{Sebastian Hidalgo}
\affil{Instituto de Astrofisica de Canarias, Via Lactea s/n 38200 La Laguna, S/C de Tenerife, 
 Spain}
\authoremail{shidalgo@ll.iac.es}

\begin{abstract}
We have obtained mosaic images of NGC 3109 in the V and I bands on 74 nights,
spanning approximately one year. From these data, we have conducted an extensive search
for Cepheid variables over the entire field of the galaxy, resulting in the discovery 
of 113 variables with periods ranging from 3.4 to 31.4 days. In this sample, 76 Cepheids,
including many long-period variables,
 were not known before. For the previously known
45 Cepheids in this galaxy, our data proved that reported periods were wrong for 14 objects;
for nearly all other previously known Cepheid variables we were able to significantly
improve on the periods. We construct period-luminosity relations from
our data and obtain reddening-corrected distance moduli of 25.72 $\pm$ 0.05 mag in V,
and 25.66 $\pm$ 0.04 mag in I. The distance modulus derived form the
reddening-independent V-I Wesenheit index turns out to be significantly
shorter (25.54 $\pm$ 0.05 mag), which indicates that in addition to the
foreground extinction of E(B-V) = 0.05 mag, there is an intrinsic to NGC
3109  redening of about 0.05 mag. Our distance obtained based on the
reddening-free Wesenheit magnitudes is consistent with earlier distance determinations
of NGC 3109 from Cepheids, and the tip of the red giant branch. 
We will improve on our distance and extinction determination  
combining our optical data with the follow-up near-infrared observations of a subsample
of NGC 3109 Cepheids. 
\end{abstract}


\section{Introduction}

In our ongoing Araucaria Project, we are improving on the usefulness of a number
of stellar distance indicators by determining their environmental dependences from a study
of these objects in a number of nearby galaxies with largely different environmental
parameters. We have described our approach in a number of previous papers (Pietrzy{\'n}ski et al. 2002, 2004;
Gieren et al. 2005a). Cepheid variables are among the most important distance indicators
we are studying. In spite of recent claims about a possible non-universality of the
period-luminosity (PL) relation (Sandage et al. 2004, Marconi et al. 2005), or a break in the relation at a period
of about 10 days (Ngeow et al. 2005, Marconi et al. 2005), previous results of the Araucaria Project 
seem to indicate that the PL relation is a superb instrument
for distance measurement, particularly in the near-infrared where reddening is not a significant
source of systematic error (Gieren et al. 2005b, 2005c; Pietrzy{\'n}ski et al. 2004, 2006).
We have been conducting wide-field optical photometric surveys for Cepheid
variables in all target galaxies of the project; so far, results were reported for NGC 300
(Pietrzy{\'n}ski et al. 2002; Gieren et al. 2004) and NGC 6822 (Pietrzy{\'n}ski et al. 2004).
We have also been carrying out near-IR (JK) follow-up observations of Cepheids in all Araucaria
target galaxies; results were reported in Gieren et al. 2005c (NGC 300), and Pietrzy{\'n}ski et al. 2006
(IC 1613). These studies have shown that a combined optical/near-infrared analysis of 
Cepheid variables in nearby, well resolved galaxies
can yield distances accurate to 3 percent.

In this paper, we are extending
our Cepheid work to the Local Group irregular galaxy NGC 3109, with the aim to provide a
sample of Cepheids with well-observed light curves in V and I bands which is near-complete
over the whole spatial extent of the galaxy, and down to a pulsation period of about 4 days.
This survey complements, and enlarges previous surveys which have covered only small areas
in NGC 3109, or which lacked the sensitivity to discover the population of short-period
and relatively faint variables. Our new Cepheid observations give us a superb database to
improve on the previous Cepheid-based distance determinations of NGC 3109 of Sandage \& Carlson (1988),
Capaccioli et al. (1992), and Musella et al. (1997), which we exploit in section 4 of
this paper.

In the first, photographic study of NGC 3109 Sandage \& Carlson (1988) had discovered 29 Cepheids. 
In this study, we recovered 28 of these 29 variables, which are among the brightest Cepheids 
in the galaxy. The Sandage \& Carlson Cepheid variable V12 was not recovered in our survey because
it fell in the gap between the two CCDs which we used to image the galaxy (Fig.
1).  Capaccioli,
Piotto and Bresolin (1992) did not discover additional Cepheids but rather corrected the B-band zero point
of Sandage's photometry and improved on the distance determination. Later Musella, Piotto and
Capaccioli (1997) obtained BVRI light curves for 24 Cepheids in a small field of NGC 3109, 16
of which were newly discovered, and bringing the total number of known Cepheid variables in the galaxy
to 45. We did not detect 6 out of 16 new Cepheids reported by Musella,
Piotto and Capaccioli (1997). Three of them, P4, P8 and P11 have periods
of about 2 days and are clearly too faint on our images to be able to
derive reliable photometry for them. The Cepheids designated as P7 and
P9 are heavily blended and we suspect that they may have been spurious
detections. Finally, P15 is not a stellar object - most probably it is a
background galaxy.
Our current survey enlarges the known Cepheid population in NGC 3109 to 113 objects and
is the first comprehensive survey of Cepheids in this galaxy which is near-complete to rather
short periods (about 4 days), and over the complete spatial extent of NGC 3109.

The paper is organized as follows. In section 2, we desribe the observations, calibrations and
data reduction procedures. In section 3, we present the catalog of Cepheid variables in NGC 3109
obtained from our observations, and in section 4 we discuss the PL relations obtained from our data,
and use them to determine the distance to NGC 3109. We discuss our results
in section 5 and summarize the main conclusions of the paper in section 6.

\section{Observations, Data Reduction and Calibration}
All the data presented in this paper were collected with the Warsaw 1.3 m 
telescope at Las Campanas Observatory. The telescope was equipped with 
a mosaic 8 chip detector, with a field of view of about 35 x 35 arcmin 
and a scale of 0.25 arcsec/pixel. For more details on the instrumental system, 
the reader is referred to the OGLE Web site (www.astrouw.edu.pl/OGLE). 
The observations were secured 
on 74 different nights during two consecutive observing seasons.
On each night, 900 sec integrations were performed through V and I filters. 
After preliminary reductions with the IRAF \footnote{IRAF is distributed by the
National Optical Astronomy Observatories, which are operated by the
Association of Universities for Research in Astronomy, Inc., under cooperative
agreement with the NSF.}  package (e.g. debiasing, flatfielding) the data were 
reduced with the OGLE III pipeline on the basis of the image-subtraction 
technique (Udalski 2003, Wo{\'z}niak 2000). 

In order to calibrate our photometry onto the standard system, about 25 Landolt 
standards were observed together with our target field during three photometric 
nights. Standard stars were observed on each of the chips, and transformation 
coefficients were derived independently for each chip, and on each of the three 
photometric nights. In order to account for small variations of the zero point in 
both filters over the mosaic field, we adopted the correction maps elaborated by 
Pietrzy{\'n}ski et al. (2004). We estimated the accuracy of the zero points 
of our photometry to be better than 0.03 mag in both filters.

Fig. 2 shows a comparison of our photometry with 
the photometry obtained by Hidalgo et al. (in preparation) with the 2.5
m Du Pont telescope at Las Campanas. It turns out that in 
both filters the zero points are consistent at the 0.03 mag level (e.g. 
within the quoted errors).

\section{The Catalog of Cepheids}
All stars observed in NGC 3109 were searched for photometric variations with periods between
0.2 and 100 days, using the analysis of variance algorithm (Schwarzenberg-Czerny 1989). 
In order to distinguish Cepheids from other types of variable stars,
we used the same criteria as in Pietrzy{\'n}ski et al. (2002). All light curves of those variables
identified as Cepheid candidates were approximated by Fourier series of order 4. We then rejected 
those objects with V amplitudes smaller than 0.4 mag, which is approximately the lower amplitude 
limit for normal classical Cepheids. For the stars passing our selection
criteria, mean V and I magnitudes were derived by integrating their light curves, which
had been previously converted onto an intensity scale, and converting the results back 
to the magnitude scale. Because of the very good quality and phase coverage of our light
curves, the statistical accuracy of the derived mean magnitudes was typically 0.01 mag
for the brighter variables, and 0.02-0.03 mag for the faintest Cepheids in our catalog.
We ended up with a list of 113 bona fide classical Cepheids with periods between 31.4 and 3.4 days.
There are evidently no truly long-period Cepheids in NGC 3109, with periods up to 50 days
or more, as in most other galaxies of the Araucaria Project we have studied so far.

Table 1 gives the individual photometric V and I observations for all Cepheid variables we found. 
The full Table 1 is available in electronic form. In
Table 2, we present their identifications, equatorial coordinates, periods, times of maximum light
in V, mean V- and I-band intensity magnitudes, and cross-identifications with previous
surveys. The accuracy of the periods can be judged from the
last digits given in column 4 of Table 2; it is typically a few units of the last digit.
We note here that for 14 Cepheids previously discovered in NGC 3109, the reported periods were
wrong. For the remaining variables with first-epoch observations, our new data helped to
improve the periods significantly.

In Fig. 3, we show phased V- and I-band light curves for some of the variables in our
catalog   which are typical for other Cepheids of similar periods, in order
to demonstrate the quality of our data.

\section{Period-Luminosity Relations and Distance Determination}

For the purpose of studying the PL relations defined by the Cepheids in NGC 3109, and determine
the distance of the galaxy, we excluded all variables with periods less than 
logP (days)=0.75 (5.6 days). This period cutoff was chosen because shortward of it
the scatter of the data points in the PL plane starts to increase significantly, as a
consequence of the increasingly lower signal-to-noise ratio in the photometry. Also, below a period of
5.6 days contamination of the sample with first-overtone pulsators is expected to become a problem
as impressively demonstrated in the LMC PL relations provided by the different microlensing
projects (e.g. Udalski et al. 1999).

Fig. 4 shows the location of all Cepheids in our catalog on the period-magnitude planes in V and I.
Stars with periods shorter than the adopted cutoff period, as well as a very few outliers,
are marked with open symbols in these plots. Our final Cepheid sample used for the analysis
of the PL relations in NGC 3109, excluding the stars marked with open circles
in Fig. 4, contains 83 objects. This subsample of Cepheids should not contain peculiar,
or strongly blended stars, and it should be free of contaminating overtone pulsators
whose inclusion in the sample
would tend to make the derived Cepheid distance to NGC 3109 too short.
In Fig. 4, an indication of a possible change of the
slope at the log(P) = 1.2 can be noticed, but it cannot be clearly 
verified due to the relatively small number of long period Cepheids in
NGC 3109. However, since the slopes of the P-L relations (see below) agree relatively well
with the corresponding slopes obtained for the LMC Cepheids
this effect should not affect our distance determination significantly. 
A plot of the location of these 83 variables on the V, V-I color-magnitude diagram 
in Fig. 5 shows that all
these stars, with the sole possible exception of one rather blue object, are indeed located in
the Cepheid instability zone, giving support to our conclusion that we
have isolated a very nearly uncontaminated sample of classical Cepheids in NGC 3109 for
our study of the PL relation, and its application for the determination of the distance to the galaxy.

In agreement with our usual procedure (e.g. Gieren et al. 2004, Pietrzy{\'n}ski et al. 2004),
we performed fits to a straight line to the 83 datapoints in the PL planes adopting for this purpose
the slopes of the LMC Cepheid PL relations established by the OGLE-II Project (Udalski 2000).
Free fits to the datapoints which do not pre-specify the slopes,
yield slopes which agree with the adopted LMC slopes within the
combined 2 $\sigma$ uncertainties, yielding no evidence for any significant departure
of the NGC 3109 PL relation slopes in V and I  from the fiducial LMC PL relations, in line
with our previous findings for NGC 300 and NGC 6822 (Gieren et al. 2004, Pietrzy{\'n}ski et al. 2004).
Fitting the OGLE LMC slopes to our data yields the following relations: \\

V = -2.775 log P + (24.418 $\pm$ 0.048) \\

I = -2.977 log P + (23.833 $\pm$ 0.042) \\

Wi = -3.000  log P + (22.908 $\pm$ 0.051) \\                                                                        
       
The linear regressions to our V, I, and Wi  Cepheid PL relations are shown in
Fig. 6 and 7. In the case of the Wi index we excluded four more widely
oustanding objects: cep033, cep057, cep078 and cep082. All of them have
either very blue or very red colors, what coused their abnormal
locations on the Wi PL relation.

Assuming the Schlegel et al. (1998) reddening law, in agreement with our previous work
in the Araucaria Project, a foreground reddening towards NGC 3109 of E(V-I) = 0.05 mag 
(Minniti, Zijlstra and Alonso 1999), and adopting a true LMC distance modulus of 18.5,
 we derive from the above zero points true distance moduli of NGC 3109 of
25.72 $\pm$ 0.05 mag in the V band, and 25.66 $\pm$ 0.04 in the I band.
We adopt the distance modulus of 25.54 $\pm$ 0.05 mag, derived from the
reddening-free Wesenheit magnitudes as our best result. As one can see
it is significantly shorter than the corresponding distance moduli obtained 
based on the V and I band magnitudes. A straightforward explanation is 
that in addition to our adopted extinction (E(B-V) = 0.05 mag), we do
have some contribution to the E(B-V) related to the intrinsic (to NGC
3109) extinction. Indeed assuming an additional intrinsic reddening 
of 0.05 mag the distances derived for the V and I band agree within 
0.02 mag with our adopted distance modulus from the Wesenheit index. 
We would like to note that we found an identical situation studying 
Cepheid P-L relations in the optical V and I bands in NGC 300 
(Gieren et al. 2004) and also 
postulated an additional intrinsic reddening as the most reasonable
 explanation. Recently Gieren et al. (2005c) measured the
total extinction in this galaxy from combined optical
and infrared data of Cepheids and indeed confirmed this suspition. 

We will precisely measure the total extinction of NGC 3109 once we will have analyzed 
follow-up near-infrared photometry of a subsample of the Cepheids in
this galaxy. 

\section{Discussion}

Our adopted Cepheid distance to NGC 3109 derived from optical photometry in the
V and I bands is 25.54 $\pm$ 0.05 mag (statistical error). This distance value
compares very well to previous distance determinations of NGC 3109 from Cepheid
variables, and the TRGB method. The most recent determination from Cepheids
is the one by Musella, Piotto and Capaccioli (1997) who analyzed BVRI photometry
of 24 Cepheids and obtained a true distance modulus of 25.67 mag. Capaccioli
et al. (1992) had obtained a distance of 25.50 $\pm$ 0.16 mag,
from B-band photographic photometry of 21 Cepheids. Lee (1993) obtained a true
distance modulus of 25.55 $\pm$ 0.1 mag from the location of the I-band TRGB
in NGC 3109, very similar to the more modern value of 25.61 $\pm$ 0.1 mag of
Minniti, Zijlstra and Alonso (1999).
Another very recent determination from the TRGB method gives 25.66 $\pm$ 0.1 mag
(Hidalgo et al., in preparation). 

As detailed in several of our previous papers in this series, for the scientific purposes
of the Araucaria Project we are interested in obtaining a set of galaxy distances
of the highest possible accuracy. We have shown (Gieren et al. 2005c, Bresolin et al. 2005,
Pietrzy{\'n}ski et al. 2006) that by combining high-quality optical and near-infrared photometry
of Cepheids, distance determinations accurate to 3 percent seem possible, minimizing systematic
problems with blending with unresolved companion stars, and most importantly with
reddening, which seems to be the most critical single systematic factor which can bias
distance determinations to nearby galaxies based on optical photometry alone. We therefore
prefer to postpone an exhaustive discussion of the various systematic uncertainties affecting
our current distance result for NGC 3109 to a follow-up paper which will present near-infrared
photometry of a subsample of the Cepheids studies in the optical spectral range in this paper,
and which will derive the ultimate, most accurate distance to NGC 3109 from its population of
Cepheid variables. The
principal value of the present paper is the presentation of a catalog of Cepheid variables
in this rather distant Local Group galaxy which increases the size of the known Cepheid
sample by almost a factor of 3, presents high-quality light curves and periods for all
these variables.

\section{Conclusions}

We have conducted a wide-field survey for Cepheid variables in the Local Group galaxy NGC 3109
which covers the entire spatial content of the galaxy, and which is complete to Cepheid periods
down to about 4 days. The survey was conducted in the V and I bands on 70 nights distributed
over an interval of approximately one year.
We have discovered 113 Cepheid variables with periods in the range from
31.4 to 3.4 days. Most of the previously known (45) Cepheids were re-discovered in our survey;
for a number of these variables with first-epoch data we were able to improve very
significantly on the periods with our data, and show that for 14 objects the previously
determined periods were wrong. We have used our data to construct period-luminosity relations
in the V and I bands which are of very good quality. Adopting a period-cutoff of log P(days) = 0.75,
and the slopes of the LMC Cepheid PL relations in V and I measured by Udalski (2000), we obtain
a reddening-corrected distance modulus of  25.54 $\pm$ 0.05 mag (statistical error) to
NGC 3109, which is in good agreement with previous distance results from optical work on
smaller Cepheid samples in this galaxy, and with determinations of the I-band magnitude
of the tip of the red giant branch. As in the former papers of the Araucaria Project,
this distance is tied to an assumed LMC distance modulus of 18.50.

We are currently engaged in analyzing a subsample of the Cepheids in NGC 3109 at near-infrared
(J and K) wavelengths. With the advent of PL relations in J and K, we expect to improve
on the accuracy of our current, optical distance determination, mainly by improving on the
value of the total reddening appropriate to NGC 3109, and to achieve a distance determination
whose accuracy is comparable to our previous work on NGC 300 (Gieren et al. 2005c) and
IC 1613 (Pietrzy{\'n}ski et al. 2006).

\acknowledgments
We would like to thank the anonymous referee for his interesting
suggestions.
WG and GP gratefully acknowledge financial support for this
work from the Chilean Center for Astrophysics FONDAP 15010003. 
Support from the DST and BW grants for Warsaw University Observatory is also acknowledged.

\begin{figure}[p] 
\vspace*{18cm}
\includegraphics{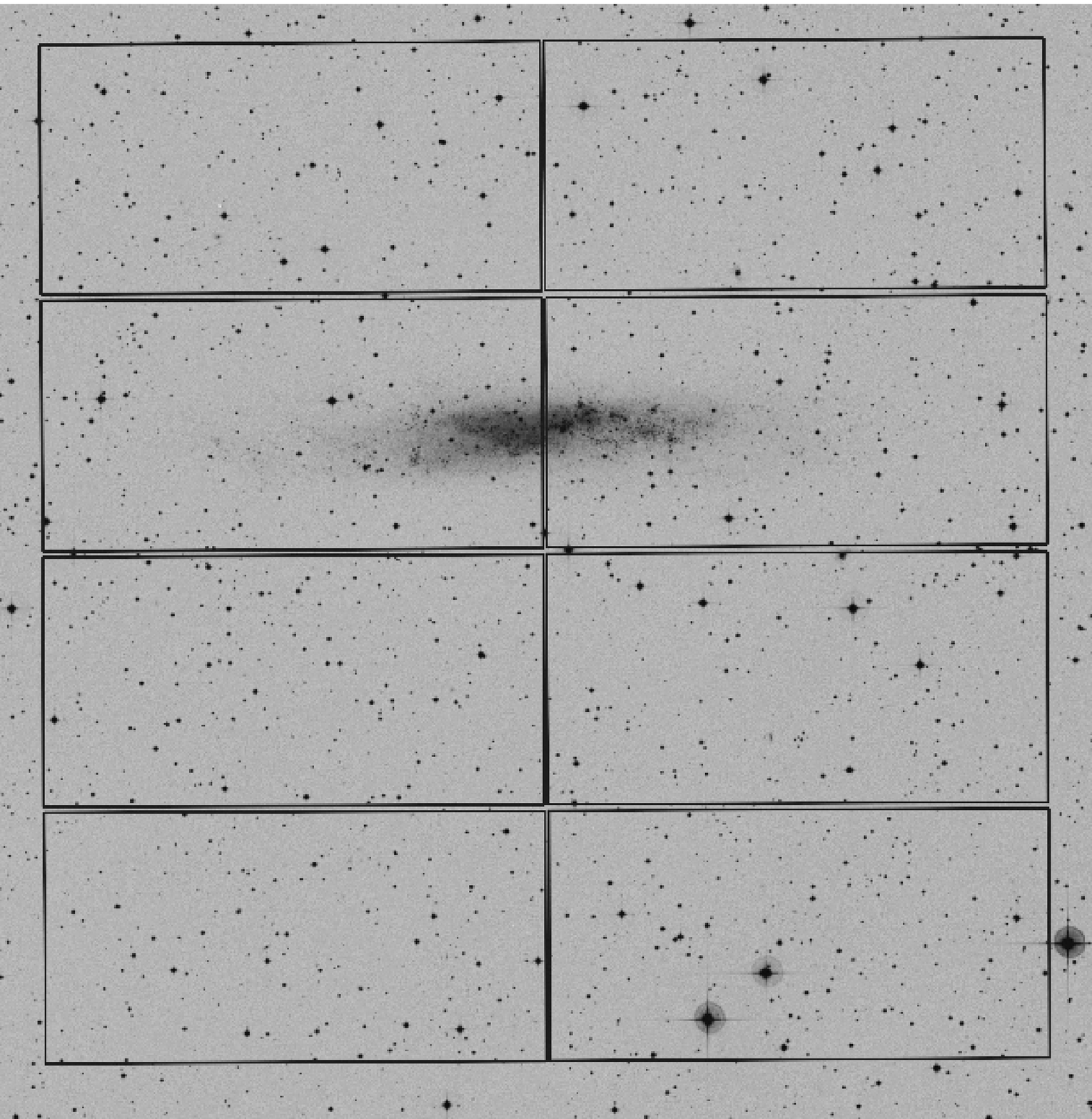} 
\caption{The location of the observed field in NGC 3109 on the DSS
blue plate. The field of view was about 35 x 35 arcmin. North is up and
East is to the left.}
\end{figure}

\begin{figure}[p]
\vspace*{18cm}
\includegraphics{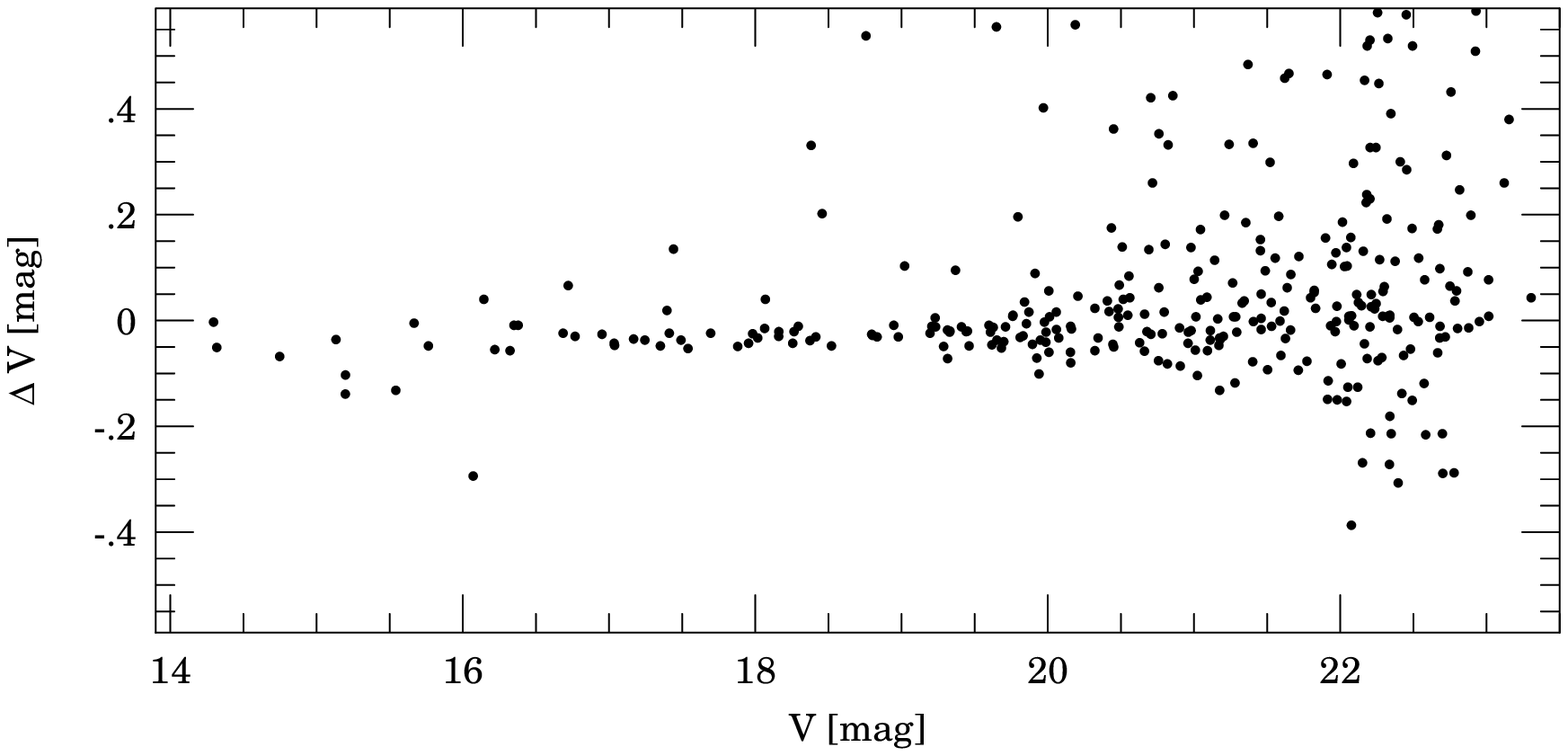}
\includegraphics{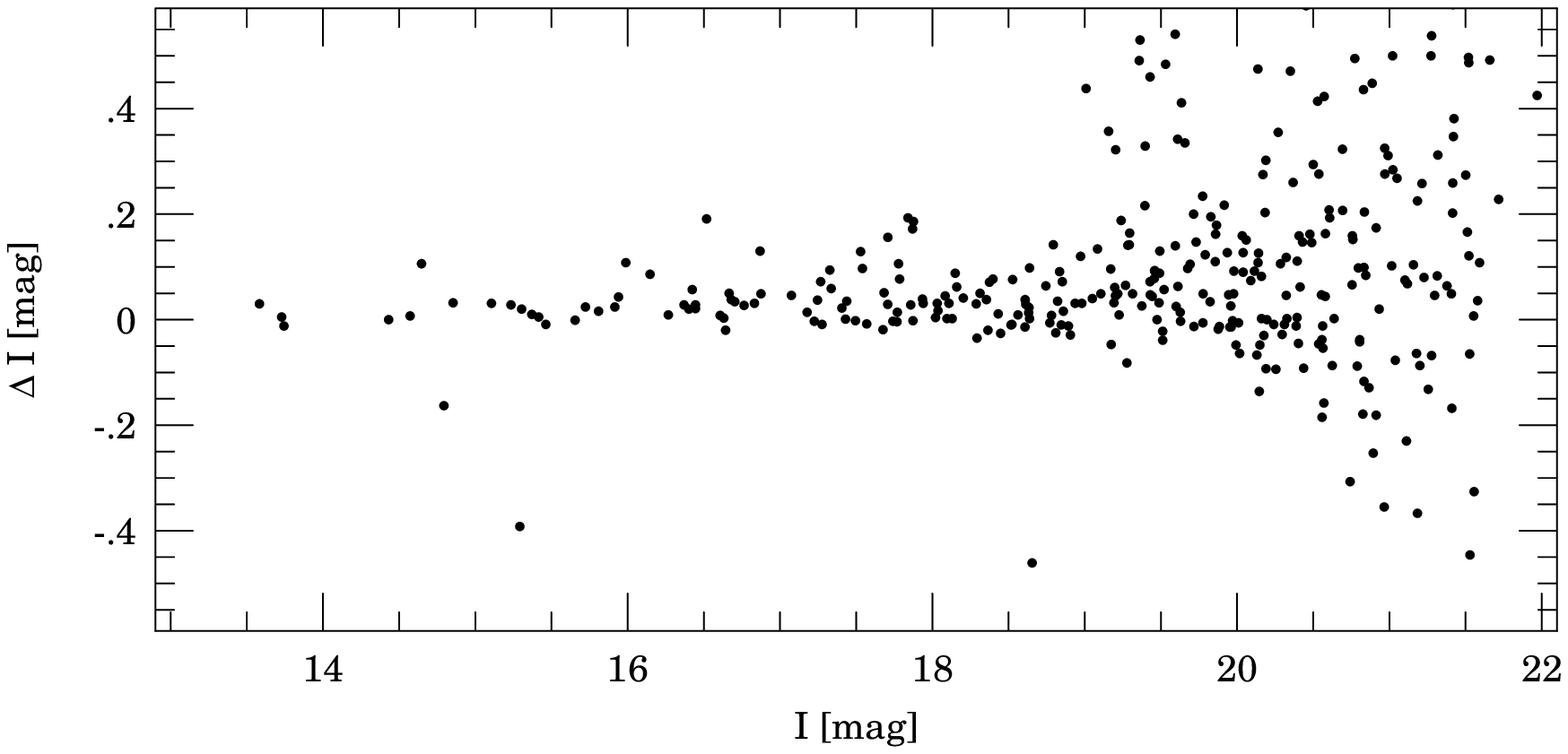}
\vspace{-3cm}
\caption{Comparison of our NGC 3109 V- and I-band photometry with the
data published by Hidalgo et al. 2006 (in preparation). The zero points
of the two studies agree to within 0.03 mag in both filters.}
\end{figure}

\begin{figure}[p]
\includegraphics{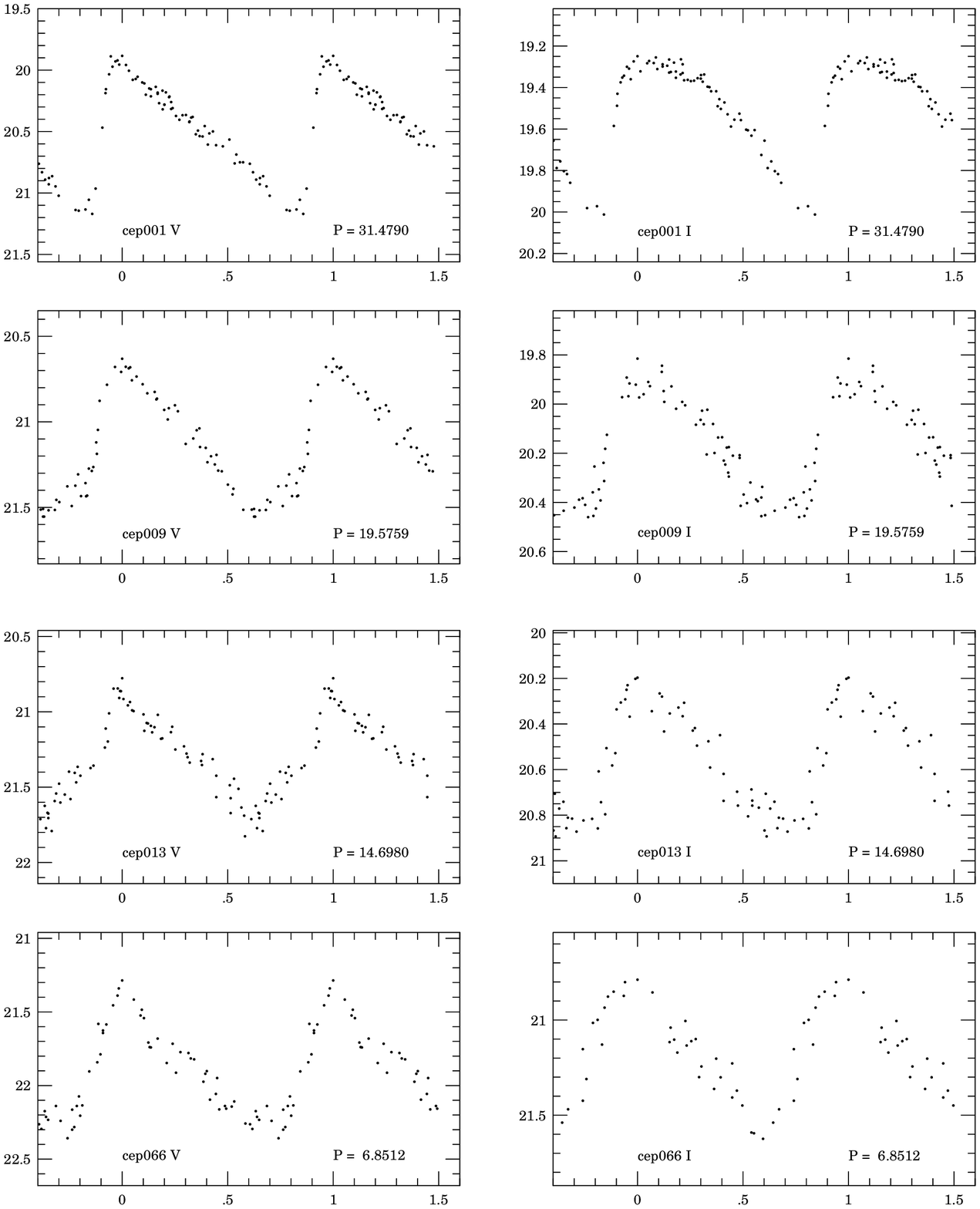}
\vspace{22cm}
\caption{V- and I-band light curves of several of the Cepheids in our catalog
from our new data. The quality of the light curves is very similar
for other variables of similar periods.}
\end{figure}

\begin{figure}[p]
\includegraphics{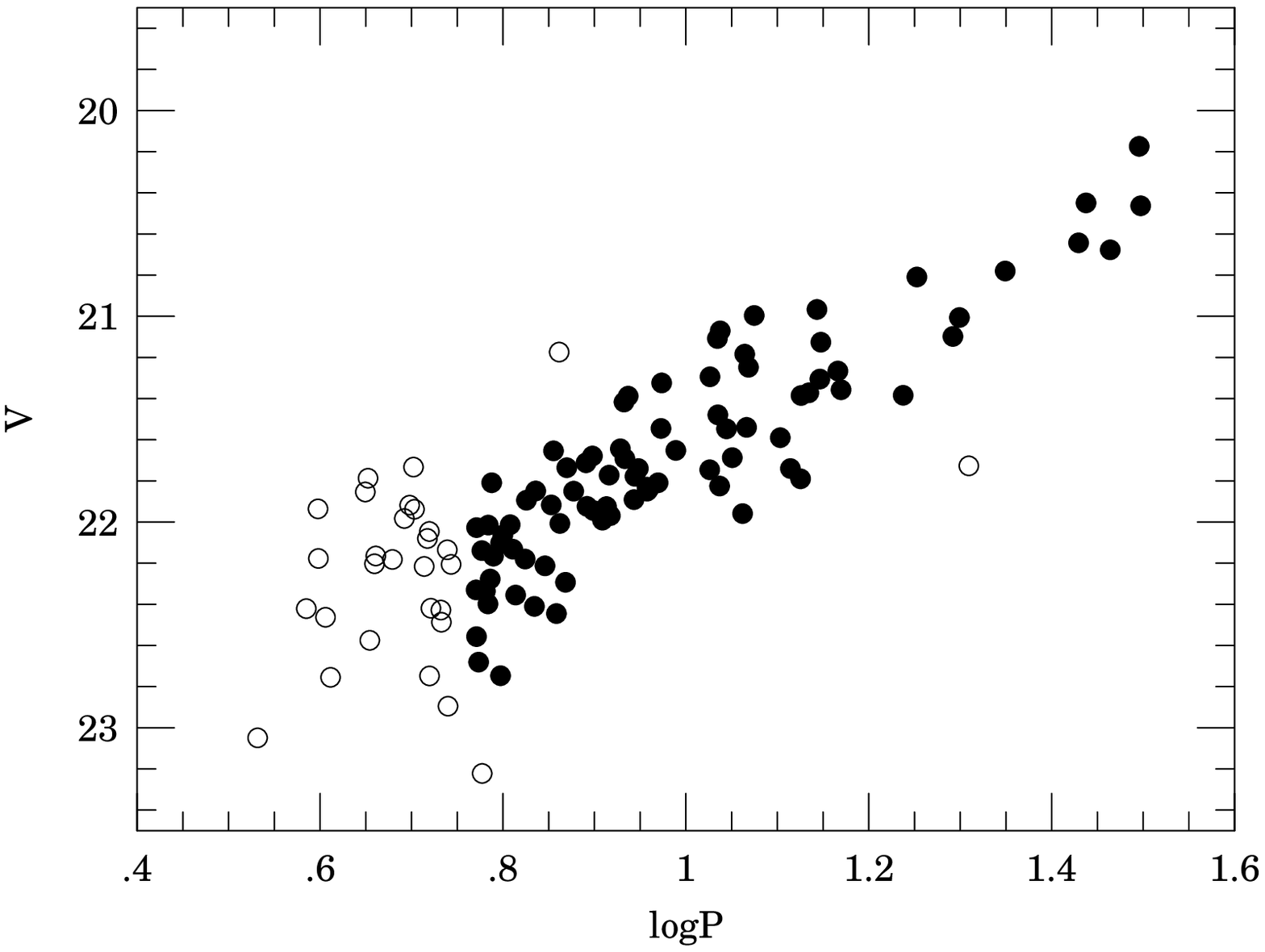}
\includegraphics{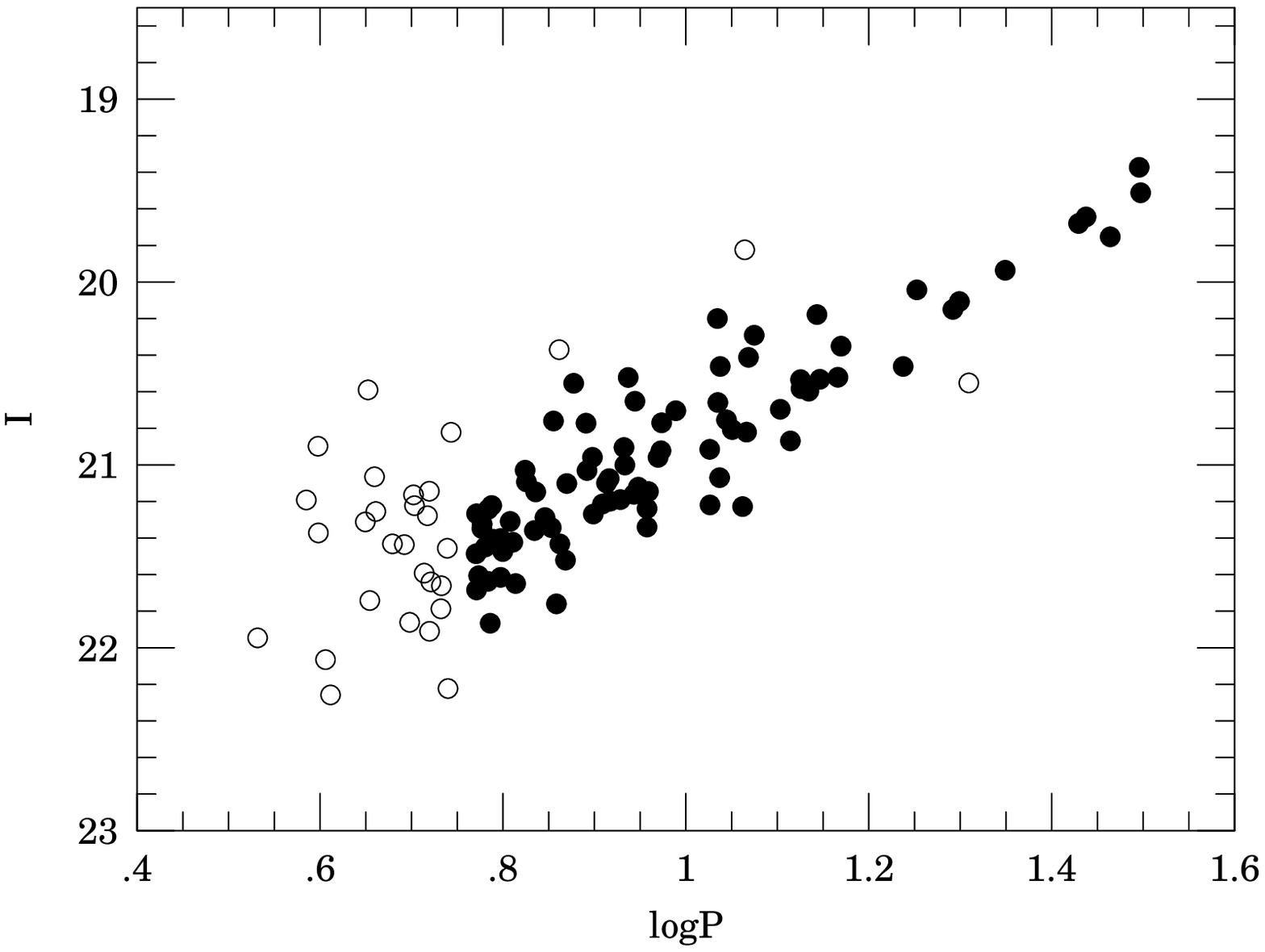}
\vspace{22cm}
\caption{Period-luminosity relations in the V and I bands, from the 113
Cepheid variables in NGC 3109 in our catalog. Cepheids with periods shorter
than our adopted period cutoff of logP (days)=0.75 (see text) were excluded
for the distance determination, as well as a few outliers. The stars not
used in the distance determination are indicated by open circles.}
\end{figure}

\begin{figure}[p]
\includegraphics{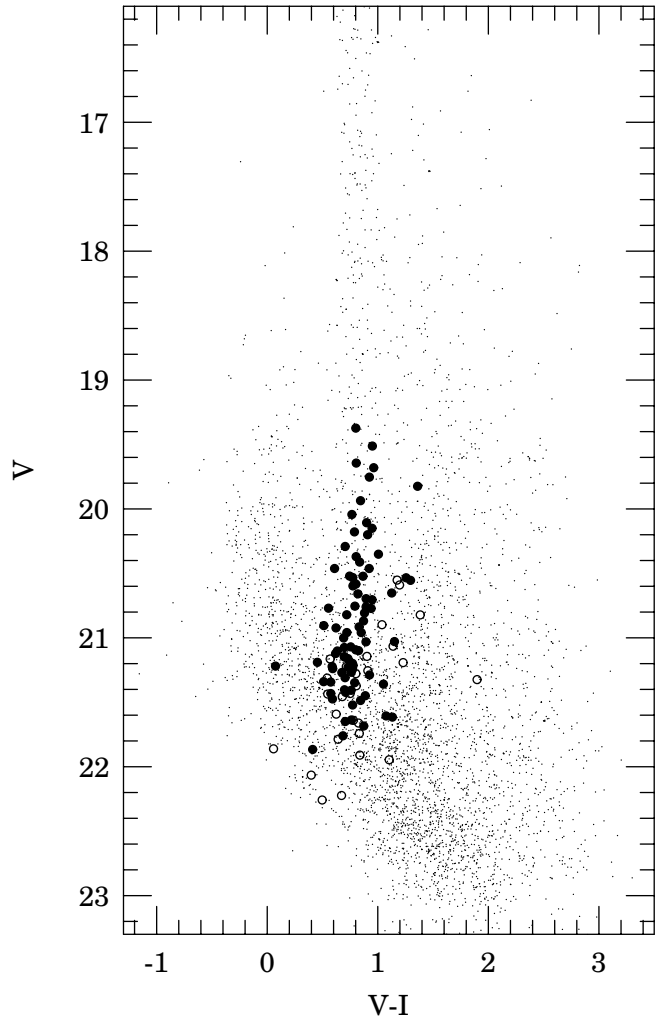}   
\vspace{18cm}
\caption{The V, V-I magnitude-color diagram for NGC 3109, showing the location
of the 113 Cepheids in our catalog. The Cepheids marked with filled circles
constitute the final sample adopted for the distance solution (see text). All these
variables lie in the instability strip for classical Cepheids.
}
\end{figure}

\begin{figure}[p]
\includegraphics{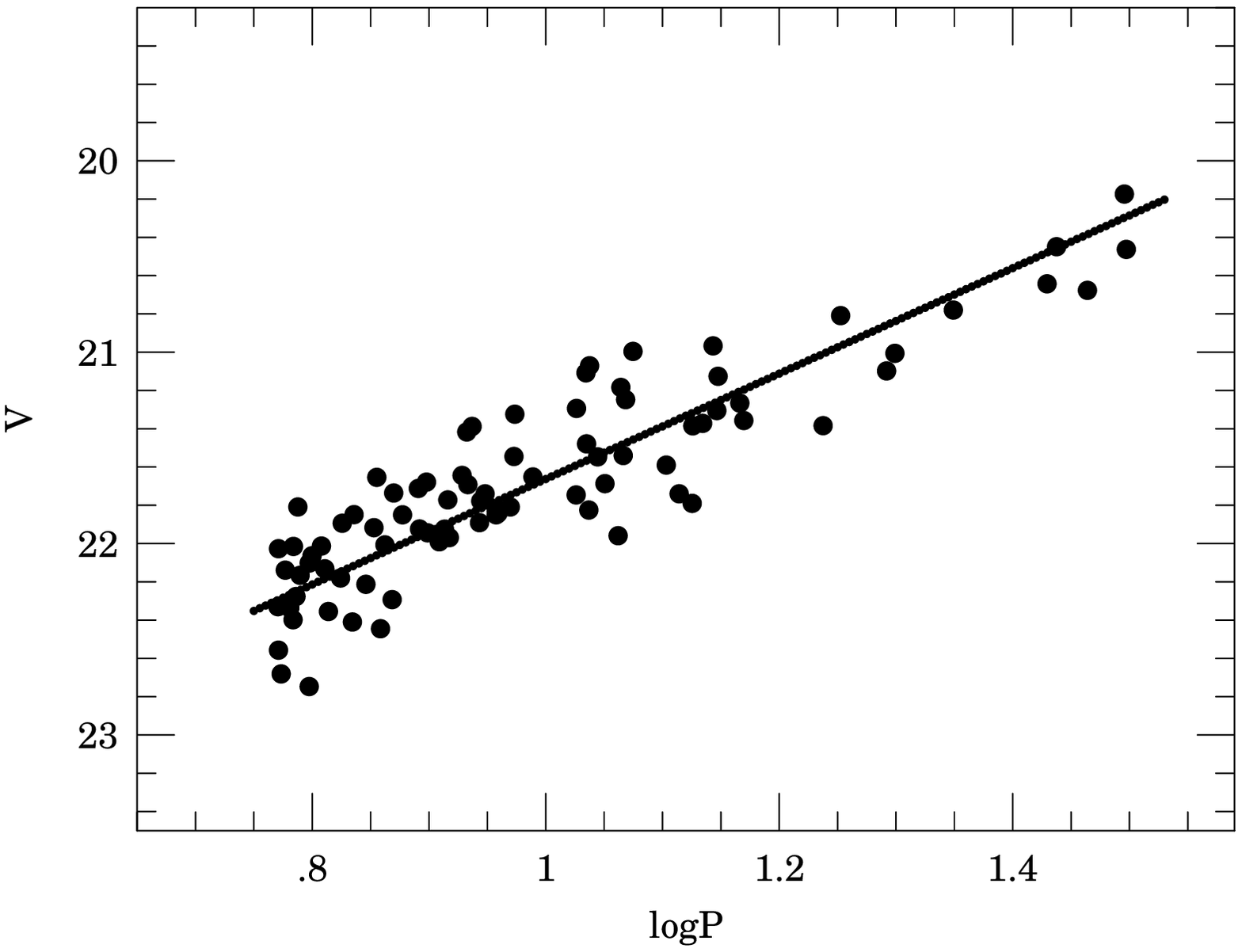}
\includegraphics{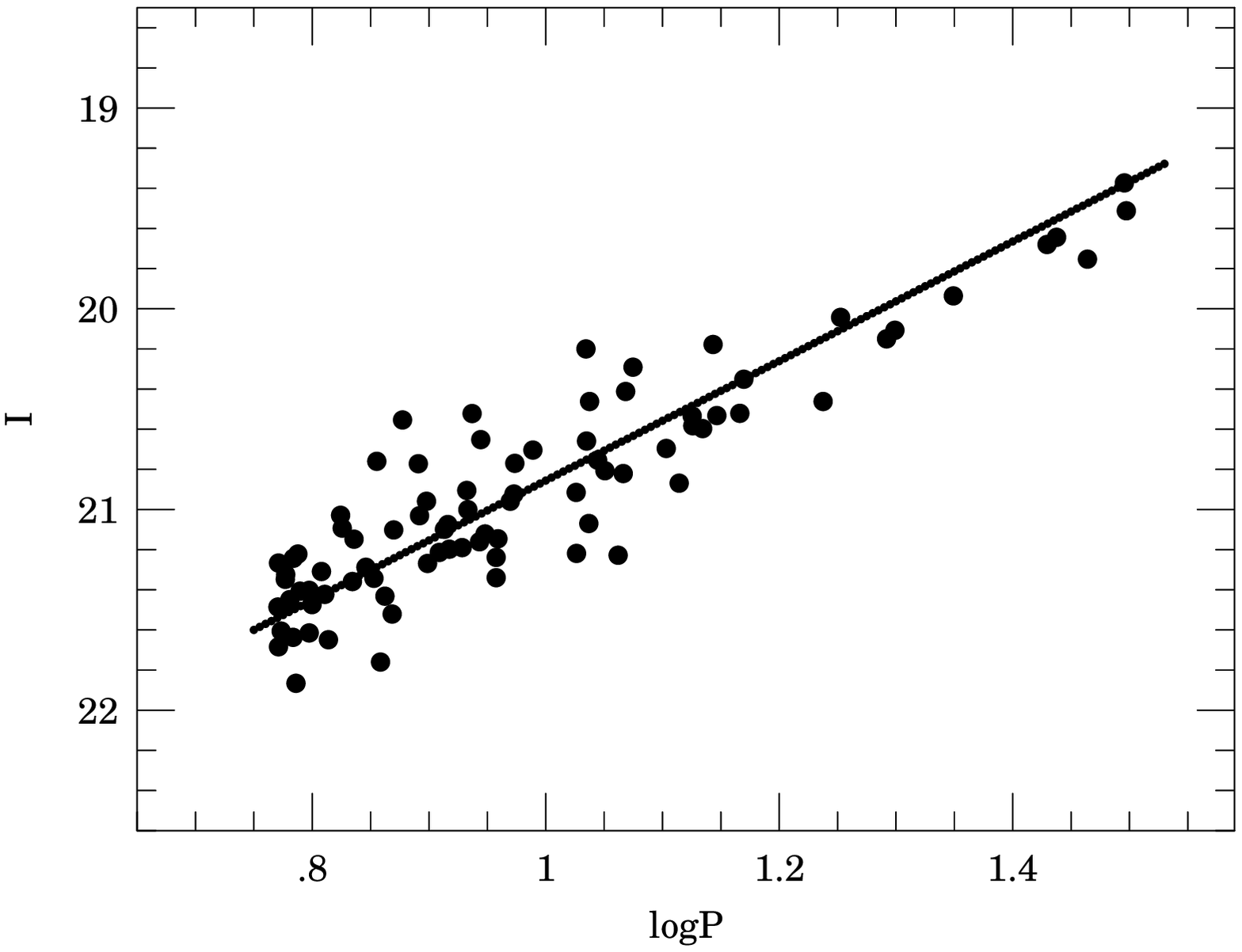}
\vspace{22cm} 
\caption{The fits to a line to our adopted Cepheid sample in V and I. The slopes of the fits
have been adopted from the LMC samples studied by the OGLE-II project.}
\end{figure}

\begin{figure}[p]
\includegraphics{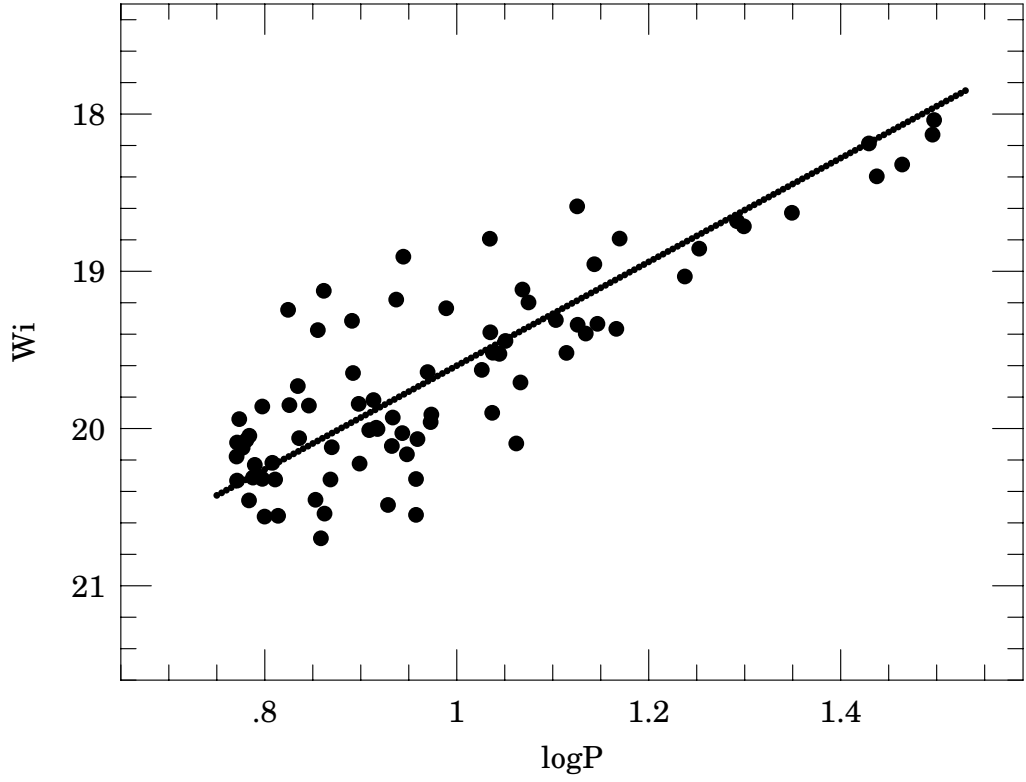}
\vspace{22cm}
\caption{Same as Fig. 6 but for the reddening-independent V-I Wesenheit 
magnitudes.}
\end{figure}

\clearpage
\begin{deluxetable}{ccccc}
\tablewidth{0pc}
\tablecaption{Journal of the Individual V and I band Observations of NGC 3109
Cepheids}
\tablehead{ \colhead{ID} & Filter & HJD & mag & $\sigma_{\rm mag}$}
\startdata
cep001 & V & 2452343.545710 &  19.929 &   0.015 \\ 
cep001 & V & 2452344.548600 &  19.884 &   0.013 \\ 
cep001 & V & 2452345.561710 &  20.005 &   0.018 \\ 
cep001 & V & 2452346.558740 &  20.072 &   0.018 \\ 
cep001 & V & 2452347.554180 &  20.100 &   0.014 \\ 
cep001 & V & 2452349.556220 &  20.136 &   0.018 \\ 
cep001 & V & 2452350.539630 &  20.167 &   0.017 \\ 
cep001 & V & 2452351.544490 &  20.220 &   0.017 \\ 
cep001 & V & 2452353.541320 &  20.367 &   0.026 \\ 
cep001 & V & 2452354.545400 &  20.423 &   0.029 \\ 
\enddata
\tablecomments{The complete version of this table is in the electronic
edition of the Journal.  The printed edition contains only
a portion of V band  data for the Cepheid variable cep001.}
\end{deluxetable}

\clearpage

\begin{deluxetable}{c c c l c c c c}
\tablecaption{Cepheids in NGC 3109}
\tablehead{
\colhead{ID} & \colhead{RA} & \colhead{DEC}  & \colhead{P} & \colhead{
${\rm
T}_{0}$} &
\colhead{$<V>$} & \colhead{$<I>$} &
\colhead{Remarks} \\
 & \colhead{(J2000)} & \colhead{(J2000)}  &
\colhead{ [days]} & \colhead{-2450000} &
\colhead{[mag]} & \colhead{[mag]} &
}
\startdata
cep001 & 10:03:21.84 & -26:09:52.9 & 31.4793 & 2337.913830 &  20.463 &  19.512& V14\\ 
cep002 & 10:03:07.83 & -26:09:54.5 & 31.270 & 2321.649470 &  20.174 &  19.373& \\ 
cep003 & 10:03:14.66 & -26:09:23.0 & 29.110 & 2337.636330 &  20.677 &  19.753& \\ 
cep004 & 10:03:19.50 & -26:10:12.8 & 27.389 & 2319.596160 &  20.449 &  19.644& \\ 
cep005 & 10:03:10.99 & -26:09:25.8 & 26.8274 & 2337.755930 &  20.643 &  19.680& V42 \\ 
cep006 & 10:02:27.01 & -26:10:13.1 & 22.3795 & 2331.567570 &  20.780 &  19.936& V80 \\ 
cep007 & 10:03:01.79 & -26:09:28.8 & 20.388 & 2331.509800 &  21.727 &  20.552& \\ 
cep008 & 10:03:43.38 & -26:10:45.4 & 19.9645 & 2343.055610 &  21.006 &  20.107& V81\\ 
cep009 & 10:02:57.57 & -26:09:55.5 & 19.5759 & 2342.458090 &  21.098 &  20.150& V64\\ 
cep010 & 10:03:28.20 & -26:10:20.5 & 17.889 & 2325.195440 &  20.809 &  20.043& \\ 
cep011 & 10:03:00.16 & -26:10:36.1 & 17.2293 & 2340.398770 &  21.384 &  20.462& V34\\ 
cep012 & 10:02:53.53 & -26:09:37.0 & 14.750 & 2332.001570 &  21.357 &  20.351& \\ 
cep013 & 10:02:46.36 & -26:09:56.9 & 14.6980 & 2336.741810 &  21.266 &  20.521& V5\\ 
cep014 & 10:03:02.97 & -26:09:13.0 & 14.062 & 2335.413040 &  21.126 &  99.999& \\ 
cep015 & 10:03:18.39 & -26:10:31.5 & 14.030 & 2337.973110 &  21.305 &  20.532& \\ 
cep016 & 10:03:10.58 & -26:11:07.6 & 13.9047 & 2335.674300 &  20.967 &  20.178& V8\\ 
cep017 & 10:02:59.18 & -26:09:29.1 & 13.6250 & 2330.266540 &  21.372 &  20.597& P3\\ 
cep018 & 10:03:10.86 & -26:09:38.6 & 13.364 & 2336.635230 &  21.385 &  20.583& \\ 
cep019 & 10:03:41.49 & -26:08:45.9 & 13.3126 & 2335.877720 &  21.790 &  20.534& V92\\ 
cep020 & 10:02:55.42 & -26:09:13.7 & 13.011 & 2341.364340 &  21.740 &  20.869& \\ 
cep021 & 10:03:05.67 & -26:08:38.5 & 12.6891 & 2340.566180 &  21.590 &  20.696& V65\\ 
cep022 & 10:02:58.65 & -26:09:11.0 & 11.864 & 2333.079460 &  20.996 &  20.291& \\ 
cep023 & 10:02:54.37 & -26:09:19.2 & 11.707 & 2333.657310 &  21.248 &  20.412& \\ 
cep024 & 10:02:52.98 & -26:08:24.5 & 11.6577 & 2335.671760 &  21.540 &  20.821& V69\\ 
cep025 & 10:02:55.00 & -26:08:56.2 & 11.596 & 2335.052600 &  21.184 &  19.824& \\ 
cep026 & 10:03:05.25 & -26:08:58.7 & 11.534 & 2338.752600 &  21.959 &  21.228& \\ 
cep027 & 10:02:58.24 & -26:08:48.2 & 11.238 & 2333.192590 &  21.687 &  20.807& \\ 
cep028 & 10:03:16.97 & -26:09:04.7 & 11.0663 & 2341.410650 &  21.547 &  20.754& V47\\ 
cep029 & 10:03:08.68 & -26:10:04.2 & 10.903 & 2339.496070 &  21.071 &  20.462& \\ 
cep030 & 10:03:11.39 & -26:09:49.2 & 10.887 & 2340.595580 &  21.825 &  21.070& \\ 
cep031 & 10:02:57.81 & -26:09:49.8 & 10.851 & 2343.875650 &  21.479 &  20.659& \\ 
cep032 & 10:03:08.44 & -26:09:40.3 & 10.826 & 2342.752810 &  21.108 &  20.200& \\ 
cep033 & 10:02:51.05 & -26:09:29.6 & 10.625 & 2336.145430 &  21.294 &  21.219& \\ 
cep034 & 10:02:43.74 & -26:09:06.0 & 10.618 & 2333.751730 &  21.746 &  20.915& \\ 
cep035 & 10:03:21.40 & -26:08:43.3 & 9.7501 & 2336.288160 &  21.652 &  20.704& \\ 
cep036 & 10:03:10.75 & -26:09:38.6 & 9.4084 & 2335.443150 &  21.324 &  20.770& \\ 
cep037 & 10:03:51.99 & -26:10:54.0 & 9.3920 & 2334.622060 &  21.545 &  20.923& \\ 
\enddata
\end{deluxetable}

\setcounter{table}{1}
\begin{deluxetable}{c c c l c c c c}
\tablecaption{Cepheids in NGC 3109 - continued}
\tablehead{
\colhead{ID} & \colhead{RA} & \colhead{DEC}  & \colhead{P} & \colhead{
${\rm T}_{0}$} & \colhead{$<V>$} & \colhead{$<I>$} & \colhead{Remarks} \\
  & \colhead{(J2000)} & \colhead{(J2000)}  &
\colhead{ [days]} & \colhead{-2450000} &
\colhead{[mag]} & \colhead{[mag]} &
}
\startdata
cep038 & 10:03:18.31 & -26:09:05.9 & 9.3242 & 2340.354910 &  21.809 &  20.959& \\ 
cep039 & 10:03:02.24 & -26:09:20.2 & 9.11136 & 2337.674660 &  21.842 &  21.146& V72\\ 
cep040 & 10:02:41.23 & -26:09:14.7 & 9.0687 & 2337.266440 &  21.831 &  21.239& \\ 
cep041 & 10:02:40.38 & -26:08:57.8 & 9.0686 & 2341.630130 &  21.850 &  21.340& \\ 
cep042 & 10:03:13.70 & -26:08:38.5 & 8.87692 & 2334.110570 &  21.740 &  21.122& V44\\ 
cep043 & 10:03:23.30 & -26:10:10.8 & 8.7970 & 2342.435330 &  21.778 &  20.652& \\ 
cep044 & 10:02:33.50 & -26:08:48.9 & 8.7726 & 2334.151600 &  21.891 &  21.161& V2\\ 
cep045 & 10:03:16.16 & -26:07:57.8 & 8.6479 & 2341.346220 &  21.388 &  20.522& \\ 
cep046 & 10:03:37.95 & -26:10:39.7 & 8.55646 & 2342.433730 &  21.692 &  21.001& V18\\ 
cep047 & 10:03:15.01 & -26:10:10.3 & 8.55708 & 2337.332320 &  21.417 &  20.905& V45\\ 
cep048 & 10:03:08.96 & -26:10:07.2 & 8.4791 & 2339.496390 &  21.644 &  21.190& \\ 
cep049 & 10:03:41.88 & -26:09:43.2 & 8.26911 & 2343.627340 &  21.969 &  21.198& V20\\ 
cep050 & 10:03:18.62 & -26:10:40.5 & 8.24330 & 2338.772320 &  21.772 &  21.075& V79\\ 
cep051 & 10:03:00.55 & -26:10:42.1 & 8.19173 & 2337.470680 &  21.925 &  21.099& V35\\ 
cep052 & 10:03:14.48 & -26:08:55.3 & 8.10451 & 2343.965380 &  21.991 &  21.214& V11\\ 
cep053 & 10:03:15.59 & -26:10:18.2 & 7.9351 & 2342.029060 &  21.944 &  21.269& V9\\ 
cep054 & 10:03:20.08 & -26:10:27.1 & 7.9038 & 2342.277150 &  21.679 &  20.959& \\ 
cep055 & 10:03:16.56 & -26:08:58.7 & 7.7960 & 2339.853750 &  21.924 &  21.031& \\ 
cep056 & 10:02:57.41 & -26:09:11.6 & 7.77376 & 2340.412360 &  21.712 &  20.772& P5\\ 
cep057 & 10:03:08.82 & -26:09:33.8 & 7.5389 & 2338.974770 &  21.850 &  20.554& \\ 
cep058 & 10:03:00.67 & -26:09:10.5 & 7.4089 & 2340.271440 &  21.736 &  21.102& \\ 
cep059 & 10:02:55.95 & -26:10:20.4 & 7.3858 & 2336.028380 &  22.293 &  21.521& \\ 
cep060 & 10:03:21.39 & -26:10:42.6 & 7.2814 & 2340.586460 &  22.007 &  21.432& \\ 
cep061 & 10:03:15.41 & -26:09:41.9 & 7.25995 & 2337.447310 &  21.174 &  20.370& V57\\ 
cep062 & 10:02:52.62 & -26:09:34.2 & 7.2186 & 2337.754860 &  22.445 &  21.760& \\ 
cep063 & 10:02:59.25 & -26:09:06.7 & 7.15940 & 2340.268610 &  21.654 &  20.760& V6\\ 
cep064 & 10:03:05.80 & -26:10:08.7 & 7.13061 & 2339.841040 &  21.917 &  21.343& V36\\ 
cep065 & 10:03:24.84 & -26:10:24.4 & 7.0137 & 2337.913530 &  22.213 &  21.288& \\ 
cep066 & 10:03:03.88 & -26:10:08.9 & 6.8511 & 2337.446440 &  21.849 &  21.148& \\ 
cep067 & 10:03:04.20 & -26:09:40.8 & 6.8294 & 2341.764590 &  22.410 &  21.359& \\ 
cep068 & 10:03:10.06 & -26:10:28.4 & 6.6932 & 2338.281130 &  21.894 &  21.093& \\ 
cep069 & 10:02:59.33 & -26:08:52.4 & 6.6735 & 2340.994290 &  22.180 &  21.029& P13\\ 
cep070 & 10:02:44.35 & -26:07:28.2 & 6.5140 & 2341.722300 &  22.355 &  21.649& \\ 
cep071 & 10:02:50.10 & -26:09:04.2 & 6.4676 & 2340.656380 &  22.132 &  21.423& \\ 
cep072 & 10:03:01.45 & -26:08:55.7 & 6.4250 & 2336.815350 &  22.013 &  21.309& P12\\ 
cep073 & 10:02:55.83 & -26:08:53.4 & 6.3065 & 2341.222110 &  22.064 &  21.474& \\ 
cep074 & 10:03:09.17 & -26:09:40.5 & 6.2719 & 2337.864370 &  22.102 &  21.403& \\ 
cep075 & 10:03:04.54 & -26:09:46.9 & 6.2713 & 2340.553600 &  22.747 &  21.615& \\ 
\enddata
\end{deluxetable}

\setcounter{table}{1}
\begin{deluxetable}{c c c l c c c c}
\tablecaption{Cepheids in NGC 3109 - continued}
\tablehead{
\colhead{ID} & \colhead{RA} & \colhead{DEC}  & \colhead{P} & \colhead{
${\rm
T}_{0}$} &
\colhead{$<V>$} & \colhead{$<I>$} &
\colhead{Remarks}  \\
 & \colhead{(J2000)} & \colhead{(J2000)}  &
\colhead{ [days]} & \colhead{-2450000} &
\colhead{[mag]} & \colhead{[mag]} &
}
\startdata
cep076 & 10:03:18.76 & -26:09:08.8 & 6.16368 & 2336.988860 &  22.166 &  21.407& V46\\ 
cep077 & 10:03:05.10 & -26:09:05.7 & 6.1329 & 2338.014500 &  21.809 &  21.222& \\ 
cep078 & 10:03:01.33 & -26:09:03.5 & 6.1098 & 2336.930330 &  22.277 &  21.866& P10\\ 
cep079 & 10:03:11.02 & -26:10:11.9 & 6.0795 & 2338.452730 &  22.015 &  21.243& \\ 
cep080 & 10:03:03.39 & -26:08:54.4 & 6.0749 & 2342.638130 &  22.398 &  21.637& \\ 
cep081 & 10:03:35.64 & -26:10:01.7 & 6.0363 & 2341.558610 &  22.336 &  21.449& \\ 
cep082 & 10:03:02.23 & -26:08:28.4 & 5.9876 & 2345.016240 &  23.222 &  21.324& \\ 
cep083 & 10:03:02.20 & -26:08:29.1 & 5.98812 & 2337.986850 &  22.139 &  21.348& V7\\ 
cep084 & 10:03:13.02 & -26:08:53.6 & 5.9343 & 2337.868200 &  22.681 &  21.606& \\ 
cep085 & 10:03:34.92 & -26:09:38.6 & 5.90434 & 2342.150450 &  22.027 &  21.267& V77\\ 
cep086 & 10:03:51.68 & -26:08:08.1 & 5.9029 & 2337.458290 &  22.557 &  21.684& \\ 
cep087 & 10:03:12.09 & -26:08:29.9 & 5.8967 & 2343.837790 &  22.330 &  21.486& \\ 
cep088 & 10:03:05.58 & -26:09:39.7 & 5.5369 & 2341.694860 &  22.206 &  20.822& \\ 
cep089 & 10:03:24.52 & -26:08:54.9 & 5.4940 & 2342.016820 &  22.895 &  22.223& \\ 
cep090 & 10:03:07.53 & -26:10:17.1 & 5.4844 & 2338.056080 &  22.135 &  21.456& \\ 
cep091 & 10:02:56.16 & -26:08:36.1 & 5.4031 & 2338.409700 &  22.487 &  21.661& \\ 
cep092 & 10:02:56.00 & -26:10:37.2 & 5.39909 & 2339.090920 &  22.428 &  21.787& P1\\ 
cep093 & 10:03:26.68 & -26:10:27.2 & 5.2624 & 2341.892040 &  22.419 &  21.640& \\ 
cep094 & 10:03:09.40 & -26:10:17.6 & 5.2442 & 2338.532080 &  22.748 &  21.910& \\ 
cep095 & 10:03:18.92 & -26:09:14.8 & 5.2420 & 2339.558630 &  22.046 &  21.144& \\ 
cep096 & 10:03:20.83 & -26:09:33.1 & 5.2163 & 2340.797600 &  22.080 &  21.278& \\ 
cep097 & 10:02:55.50 & -26:10:05.5 & 5.17264 & 2342.796630 &  22.216 &  21.592& P2\\ 
cep098 & 10:03:17.06 & -26:09:05.8 & 5.0493 & 2343.287550 &  21.938 &  21.223& \\ 
cep099 & 10:02:58.38 & -26:09:10.5 & 5.03650 & 2339.717280 &  21.733 &  21.164& P6\\ 
cep100 & 10:02:58.59 & -26:09:49.7 & 4.9883 & 2341.693950 &  21.918 &  21.861& \\ 
cep101 & 10:03:05.41 & -26:08:59.6 & 4.9221 & 2337.756730 &  21.983 &  21.436& \\ 
cep102 & 10:03:15.00 & -26:08:59.1 & 4.7765 & 2342.741240 &  22.182 &  21.432& \\ 
cep103 & 10:02:54.09 & -26:08:57.7 & 4.5808 & 2339.387750 &  22.165 &  21.255& \\ 
cep104 & 10:02:53.62 & -26:09:54.1 & 4.5658 & 2341.226270 &  22.203 &  21.065& \\ 
cep105 & 10:03:14.83 & -26:10:03.1 & 4.5120 & 2340.292010 &  22.575 &  21.742& \\ 
cep106 & 10:03:08.03 & -26:08:52.2 & 4.4919 & 2341.467330 &  21.787 &  20.590& \\ 
cep107 & 10:02:44.12 & -26:09:07.4 & 4.4613 & 2343.882180 &  21.854 &  21.312& \\ 
cep108 & 10:02:55.76 & -26:09:34.4 & 4.0877 & 2341.161440 &  22.755 &  22.258& \\ 
cep109 & 10:03:46.95 & -26:10:25.0 & 4.036 & 2344.285480 &  22.463 &  22.065& \\ 
cep110 & 10:03:18.92 & -26:09:10.7 & 3.9641 & 2341.545940 &  22.177 &  21.372& \\ 
cep111 & 10:03:02.45 & -26:08:32.8 & 3.9604 & 2343.848810 &  21.936 &  20.898& \\ 
cep112 & 10:02:46.74 & -26:09:33.6 & 3.8455 & 2341.021000 &  22.421 &  21.192& \\ 
cep113 & 10:02:59.50 & -26:08:36.7 & 3.4016 & 2340.045330 &  23.049 &  21.946& P14\\ 
\enddata
\tablecomments{Symbols V and P correspond to the designations used in
the work by Sandage and Carlson (1988) and Musella, Piotto and
Capaccioli (1997), respectively.}
\end{deluxetable}


\begin{references}
\reference{} Bresolin, F., Pietrzy{\'n}ski, G., Gieren, W. and Kudritzki, R.P., 2005,
\apj, 634, 1020

\reference{} Capaccioli, M., Piotto, G., and Bresolin, F., 1992, \aj, 103, 1151

\reference{} Gieren, W., Pietrzy{\'n}ski, G., Walker, A., Bresolin, F., Minniti, D.,
Kudritzki, R.-P., Udalski, A., Soszy{\'n}ski, I., Fouqu{\'e}, P., Storm, J.,
and Bono, G., 2004, \aj, 128, 1167

\reference{} Gieren, W., Pietrzy{\'n}ski, G., Bresolin, F., Kudritzki, R.-P., Minniti, D.,
Urbaneja, M., Soszy{\'n}ski, I., Storm, J., Fouqu{\'e}, P., Bono, G., Walker, A.R.,
and Garc{\'i}a, J.A., 2005a, The ESO Messenger, 121, 23 

\reference{} Gieren, W., Storm, J., Barnes III, T.G., Fouqu{\'e}, P., Pietrzy{\'n}ski, G.,
and Kienzle, F., 2005b, \apj, 627, 224

\reference{} Gieren, W., Pietrzy{\'n}ski, G., Soszy{\'n}ski, I., Bresolin, F., Kudritzki, R.P.,
Minniti, D. and Storm, J., 2005c, \apj, 628, 695

\reference{} Lee, M.G., 1993, \apj, 408, 409

\reference{} Marconi, M., Musella, I., Fiorentino, G., 2005, \apj, 632, 590

\reference{} Minniti, D., Zijlstra, A.A., and Alonso, M.V., 1999, \aj, 117, 881

\reference{} Musella, I., Piotto, G., and Capaccioli, M., 1997, \aj, 114, 976

\reference{} Ngeow, Ch.Ch., Kanbur, S.M., Nikolaev, S., Buonaccorsi, J.,
Cook, K.H., Welch, D.L., 2005, \mnras, 363, 831 

\reference{} Pietrzy{\'n}ski, G., Gieren, W., Fouqu{\'e}, P., and Pont, F.,
 2002, \aj, 123, 789

\reference{} Pietrzy{\'n}ski, G., Gieren, W., Udalski, A., Bresolin, F., Kudritzki, R.P.,
Soszy{\'n}ski, I., Szyma{\'n}ski, M. and Kubiak, M., 2004, \aj, 128, 2815

\reference{} Pietrzy{\'n}ski, G., Gieren, W., Soszy{\'n}ski, I., Bresolin, F., Kudritzki, R.-P.,
Dall'Ora, M., Storm, J., and Bono, G., 2006, \apj, 642, 216

\reference{} Sandage, A., and Carlson, G., 1988, \aj, 96, 1599

\reference{} Sandage, A., Tammann, G.A., and Reindl, B., 2004, A\&A, 424, 43

\reference{} Schlegel, D.J., Finkbeiner, D.P., and Davis, M., 1998, \apj, 500, 
525

\reference{} Schwarzenberg-Czerny, A., 1989, \mnras, 241, 153

\reference{plc} Udalski, A., Szyma\'nski, M., Kubiak, M., Pietrzy\'nski, G.,
Soszy{\'n}ski, I.,
Wo\'zniak, P., and \.Zebru\'n, K., 1999, Acta Astron., 49, 201

\reference{} Udalski, A., 2000, Acta Astron., 50, 279

\reference{} Udalski, A., 2003, Acta Astron., 53, 291

\reference{} Wo{\'z}niak, P.,  2000, Acta Astron., 50, 421

\end{references}
\end{document}